\documentclass[twoside,twocolumn,english,preprintnumbers,amssymb,amsmath,aps,floatfix,prl,nofootinbib,superscriptaddress]{revtex4}
\usepackage[T1]{fontenc}
\usepackage[latin9]{inputenc}
\usepackage{color}
\usepackage{babel}
\usepackage{bm}
\usepackage{amsbsy}
\usepackage{amstext}
\usepackage{amssymb}
\usepackage{graphicx}
\usepackage{esint}
\usepackage[unicode=true,pdfusetitle,
 bookmarks=true,bookmarksnumbered=false,bookmarksopen=false,
 breaklinks=false,pdfborder={0 0 1},backref=false,colorlinks=true]
 {hyperref}
\hypersetup{
 pdfborderstyle=,linkcolor=blue,citecolor=cyan}

\makeatletter
\@ifundefined{textcolor}{}
{%
 \definecolor{BLACK}{gray}{0}
 \definecolor{WHITE}{gray}{1}
 \definecolor{RED}{rgb}{1,0,0}
 \definecolor{GREEN}{rgb}{0,1,0}
 \definecolor{BLUE}{rgb}{0,0,1}
 \definecolor{CYAN}{cmyk}{1,0,0,0}
 \definecolor{MAGENTA}{cmyk}{0,1,0,0}
 \definecolor{YELLOW}{cmyk}{0,0,1,0}
}

\usepackage{babel}

\usepackage{xcolor}

\makeatother

\begin{document}
\title{Probing vortical structures in heavy-ion collisions at
RHIC-BES energies \\ through helicity polarization}
\author{Cong Yi}
\email{congyi@mail.ustc.edu.cn}

\affiliation{Department of Modern Physics, University of Science and Technology
of China, Hefei, Anhui 230026}
\author{Xiang-Yu Wu}
\email{xiangyuwu@mails.ccnu.edu.cn}

\affiliation{Institute of Particle Physics and Key Laboratory of Quark and Lepton
Physics (MOE), Central China Normal University, Wuhan, Hubei, 430079}
\author{Di-Lun Yang}
\email{dlyang@gate.sinica.edu.tw}

\affiliation{Institute of Physics, Academia Sinica, Taipei, 11529}
\author{Jian-Hua Gao}
\email{gaojh@sdu.edu.cn}

\affiliation{Shandong Provincial Key Laboratory of Optical Astronomy and Solar-Terrestrial
Environment, Institute of Space Sciences, Shandong University, Weihai,
Shandong 264209}
\author{Shi Pu}
\email{shipu@ustc.edu.cn}

\affiliation{Department of Modern Physics, University of Science and Technology
of China, Hefei, Anhui 230026}
\author{Guang-You Qin}
\email{guangyou.qin@ccnu.edu.cn}

\affiliation{Institute of Particle Physics and Key Laboratory of Quark and Lepton
Physics (MOE), Central China Normal University, Wuhan, Hubei, 430079}

\begin{abstract}
We investigate the hydrodynamic helicity polarization of $\Lambda$
hyperons, defined as the projection of the spin polarization vector along the directions of particle momenta, at RHIC-BES energies by utilizing the relativistic (3+1)D
CLVisc hydrodynamics framework with SMASH initial conditions. 
As opposed to local spin polarization at high energy collisions, our hydrodynamic simulations demonstrate that the helicity polarization induced by the kinetic vorticity dominates over other contributions at intermediate and low collision energies. Our findings provide an opportunity to probe the fine structure of local kinetic vorticity as a function of azimuthal angle at intermediate and low collision energies by mapping our predictions to the future measurements in experiments. 
\end{abstract}
\maketitle

\section{Introduction}





Spin, as a fundamental property of particles, plays a critical role
in high-energy physics, e.g. proton spin puzzles (see recent reviews
\citep{Leader:2013jra,Wakamatsu:2014zza} and references therein).
Recently, a major breakthrough related to the spin polarization in
the relativistic heavy ion collisions has drawn widespread attentions.
In non-central heavy-ion collisions, two heavy nuclei are accelerated
to nearly the speed of light and collide with each other. These collisions
generates a large amount of orbital angular momentum, estimated to
be on the order of $10^{5}\hbar$. Such huge orbital angular momentum
will be partially converted into spin polarization of the hadrons
by spin-orbit coupling proposed by the pioneer works \citep{Liang:2004ph,Liang:2004xn,Gao:2007bc}.
The STAR collaboration at the Relativistic Heavy-Ion Collider (RHIC)
has measured the global polarization of $\Lambda$ and $\overline{\Lambda}$
hyperons \citep{STAR:2017ckg}. The results show that the vorticity
of the quark-gluon plasma (QGP) generated in the collisions is as
large as $\omega\sim10^{22}~s^{-1}$, making it the fastest vortical
system observed in nature to date. The global polarization has been
well-understood by various phenomenological models\citep{Becattini:2007nd,Becattini:2007sr,Becattini:2013fla,Fang:2016vpj,Karpenko:2016jyx,Xie:2017upb,Li:2017slc,Sun:2017xhx,Shi:2017wpk,Wei:2018zfb,Xia:2018tes,Shi:2019wzi,Fu:2020oxj,Ryu:2021lnx,Lei:2021mvp,Vitiuk:2019rfv,Wu:2022mkr}.

Interestingly, the polarization along the beam and out-of-plane directions,
namely the local spin polarization has been measured by STAR \citep{Niida:2018hfw,Adam:2019srw}
and ALICE \citep{ALICE:2021pzu}, and are investigated by many different
models, e.g. the statistical methods \citep{Becattini:2017gcx,Becattini:2022zvf,Palermo:2023cup},
quantum kinetic theory \citep{Gao:2019znl,Weickgenannt:2019dks,Weickgenannt:2020aaf,Hattori:2019ahi,Wang:2019moi,Yang:2020hri,Weickgenannt:2020sit,Li:2019qkf,Liu:2020flb,Weickgenannt:2021cuo,Wang:2021qnt,Sheng:2021kfc,Huang:2020wrr,Wang:2020dws,Stephanov:2012ki,Son:2012zy,Gao:2012ix,Chen:2012ca,Manuel:2013zaa,Manuel:2014dza,Chen:2014cla,Chen:2015gta,Hidaka:2016yjf,Hidaka:2017auj,Mueller:2017lzw,Hidaka:2018ekt,Hidaka:2018mel,Gao:2018wmr,Huang:2018wdl,Liu:2018xip,Lin:2019ytz,Lin:2019fqo,Yamamoto:2020zrs,Gao:2020vbh, Gao:2020pfu, Hidaka:2022dmn,Fang:2022ttm},
spin hydrodynamics \citep{Hattori:2019lfp,Fukushima:2020qta,Fukushima:2020ucl,Li:2020eon,She:2021lhe,Montenegro:2017lvf,Montenegro:2017rbu,Florkowski:2017ruc,Florkowski:2018myy,Becattini:2018duy,Florkowski:2018fap,Yang:2018lew,Bhadury:2020puc,Shi:2020qrx,Gallegos:2021bzp,Hongo:2021ona,Florkowski:2017dyn,Florkowski:2018ahw,Florkowski:2019qdp,Florkowski:2019voj,Bhadury:2020cop,Shi:2020htn,Singh:2020rht,Wang:2021ngp,Liu:2020ymh,Peng:2021ago,Florkowski:2021wvk,Copinger:2022jgg,Wang:2021wqq},
other effective theories \citep{Zhang:2019xya,Liu:2020dxg,Liu:2021uhn}
and phenomenological simulations \citep{Xie:2017upb,Xia:2018tes,Voloshin:2017kqp,Liu:2019krs,Wei:2018zfb,Wu:2020yiz,Wu:2019eyi,Fu:2020oxj,Xia:2019fjf,Becattini:2019ntv,Li:2021jvn,Florkowski:2021xvy,Sun:2021nsg}.
It is found that the local spin polarization can be induced by various
sources, including the thermal vorticity, shear viscous tensor, fluid
acceleration, gradient of baryon chemical potential over temperature,
and electromagnetic fields \citep{Hidaka:2017auj,Liu:2020dxg,Liu:2021uhn,Becattini:2021suc,Fu:2021pok,Becattini:2021iol,Yi:2021ryh,Ryu:2021lnx,Alzhrani:2022dpi,Buzzegoli:2022fxu,Palermo:2022lvh}.
These studies have also been extended to low-energy collisions \citep{Aziz:2021HADES,STAR:2021beb,Ivanov:2019ern,Deng:2020ygd,Guo:2021udq,Ayala:2021xrn,Deng:2021miw,Fu:2020oxj,Ryu:2021lnx,Wu:2022mkr,Fu:2022myl,Fu:2022oup} and
isobaric collisions \citep{STAR:2023eck}, as well as to the discussion
on the vortical smoke rings \citep{Lisa:2021zkj,Serenone:2021zef}.
Despite the global polarization has provided insight into the kinetic
or thermal vorticity as a function of collision energies, the fine
structure of the vorticity, such as its dependence on the azimuthal
angle, have not been fully explored. These information may not been
accurately captured by the local spin polarization due to the considerable
influence of other sources beyond the thermal vorticity. In this work,
we extend our previous studies \citep{Yi:2021unq} and demonstrate
that the helicity polarization can help us to probe the fine structure
of kinetic vorticity in low-energy collisions. 


Helicity polarization, defined as the projection
of spin polarization vector onto the direction of the particles' momentum,
is widely used for characterizing the spin polarization
in high energy physics \citep{leader_2001}. In many high-energy scattering processes,
there are no preferred quantization directions for spin, unlike the
case of global polarization where the direction of the initial orbital
angular momentum is naturally chosen as the quantization direction.
In such cases, helicity is often preferred over spin to describe the
spin polarization. 


Back to the heavy ion collisions, the use of helicity polarization
allows us to probe a distinct feature of the spin polarization for
$\Lambda$ and $\overline{\Lambda}$ hyperons. As mentioned previously,
it is challenging to distinguish the local polarization of $\Lambda$
hyperons induced by different sources through the experimental measurements.
Remarkably, our previous study \citep{Yi:2021unq} has found that
the local helicity polarization induced by thermal vorticity dominates
over other contributions at $\sqrt{s_{NN}}=200$ GeV Au+Au collisions.
We argue that helicity polarization induced by kinetic vorticity,
as part of thermal vorticity, will play the crucial role to the total
helicity polarization in the low-energy collisions. Consequently,
we further proposed that this finding can be utilized to probe the
local strength of kinetic vorticity as a function of azimuthal angle
in low-energy collisions by measuring helicity polarization. 

To verify our conjecture, we utilize the relativistic (3+1)D CLVisc
hydrodynamics framework \citep{Pang:2012he, Wu:2021fjf, Wu:2022mkr}
to investigate the azimuthal angle dependence of hydrodynamic helicity
polarization at RHIC-Beam Energy Scan (BES) energies in this work.
We report the numerical simulation of hydrodynamic helicity polarization
at $\sqrt{s_{NN}}=7.7,19.6,39\text{ GeV}$ Au+Au collisions with Simulating
Many Accelerated Strongly-interacting Hadrons (SMASH) \citep{Weil:2016zrk, Schafer:2019edr, Mohs:2019iee, Hammelmann:2019vwd, Mohs:2020awg, Schafer:2021csj, Inghirami:2022afu}
initial condition. As anticipated, the helicity polarization induced
by the kinetic vorticity is one order of magnitude larger than other
contributions in low-energy collisions. This finding holds even when
we choose A-Multi-Phase-Transport (AMPT) \citep{Lin:2004en, Wu:2021hkv, Wu:2018cpc, Zhao:2017yhj}
initial conditions or different baryon diffusion coefficients. 
Our study presents a novel approach to investigate the structure of
kinetic vorticity in low-energy heavy-ion collisions by connecting
hydrodynamic simulations with the measurable helicity polarization.
The finding can also provide a baseline for the investigation on local
parity violation through the correlations of helicity polarization
proposed by \citep{Becattini:2020xbh,Gao:2021rom}.

This paper is organized as follows. In Sec.~\ref{sec:Helicity-polarization},
we briefly introduce the theoretical framework and hydrodynamical setup
for the helicity polarization. We present our numerical results of
the helicity polarization at various collision energies, initial conditions,
and baryon diffusion coefficients in Sec.~\ref{sec:Numerical-results-from} and summarize our findings in Sec.~\ref{sec:Conclusion-and-discussion}.
Throughout this work, we adopt the Minkowski metric $g_{\mu\nu}=\textrm{diag}\{+,-,-,-\}$
and the projector $\Delta^{\mu\nu}=g^{\mu\nu}-u^{\mu}u^{\nu}$ with
$u^{\mu}$ being fluid velocity.

\begin{figure*}[thb]
\includegraphics[scale=0.35]{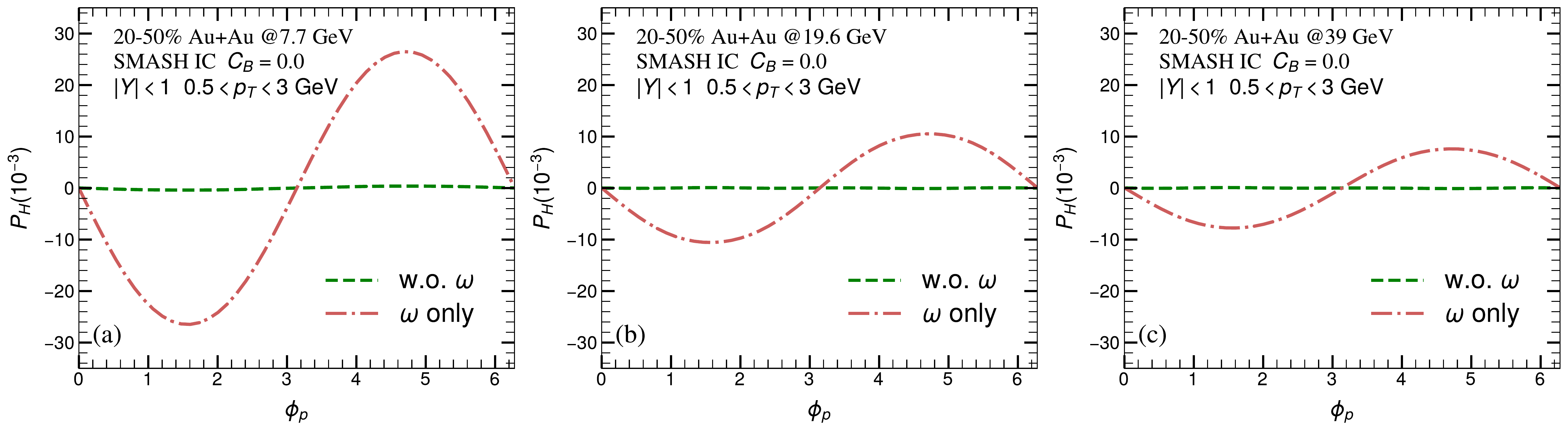}
\caption{Helicity polarization as a function of azimuthal angle $\phi_{p}$
in $20-50\%$ centrality at $\sqrt{s_{NN}}=7.7,19.6,39$ GeV Au+Au
collisions with SMASH initial condition. The red dashed-dotted lines
denote the helicity polarization induced by the kinetic vorticity
$\omega$ only, i.e. $P_{H}^{\omega}$. The green dashed lines
represent the total helicity polarization excluding $P_{H}^{\omega}$.
\label{fig:cb0}}
\end{figure*}

\section{Theoretical and numerical framework \label{sec:Helicity-polarization}}

We follow Refs. \citep{Becattini:2017gcx, Becattini:2020xbh, Gao:2021rom,Yi:2021unq}
to introduce the theoretical framework for the helicity polarization
in relativistic heavy ion collisions. The helicity polarization for a relativistic particle with mass $m$ is
defined as
\begin{eqnarray}
S^{h} & = & \text{\ensuremath{\widehat{\mathbf{p}}}\ensuremath{\cdot}}\mathcal{\boldsymbol{S}}(\mathbf{p}).\label{eq:def_Sh}
\end{eqnarray}
Here, we parameterize the momentum of an on-shell particle as $p^{\mu}=(\sqrt{|\mathbf{p}|^{2}+m^{2}},\mathbf{p})=(\sqrt{p_{T}^{2}+m^{2}}\cosh Y,p_{T}\cos\phi_{p},p_{T}\sin\phi_{p},\sqrt{p_{T}^{2}+m^{2}}\sinh Y)$,
where $p_{T}$ is the transverse momentum, $Y$ is the momentum rapidity
and $\phi_{p}$ is the azimuthal angle, $\mathbf{\ensuremath{\widehat{\mathbf{p}}}=}\mathbf{p}/|\mathbf{p}|$
is the unit vector along the direction of momentum, and $\mathcal{\boldsymbol{S}}(\mathbf{p})$
is the spatial component of the single-particle mean spin polarization
vector $S^{\mu}(p)$. The spin polarization vector $S^{\mu}(p)$ for
fermionic systems can be evaluated by using the modified Cooper-Frye
formula \citep{Becattini:2013fla, Fang:2016vpj,Hidaka:2018mel,  Yi:2021ryh},
under the assumption of local thermal equilibrium,
\begin{eqnarray}
\mathcal{S}^{\mu}(\mathrm{p}) & = & \frac{\int d\Sigma\cdot p\mathcal{J}_{5}^{\mu}(p,X)}{2m_{\Lambda}\int d\Sigma\cdot\mathcal{N}(p,X)},\label{eq:Cooper-Frye formula}
\end{eqnarray}
where $d\Sigma^{\mu}$ is the normal vector of the freeze-out hyper-surface,
$m_{\Lambda}$ denotes the mass of $\Lambda$ hyperons, and $\mathcal{J}_{5}^{\mu}(p,X)$
and $\mathcal{N}^{\mu}(p,X)$ stand for the axial-charge and number-density 
current in the phase space, respectively. 

Inserting the $\mathcal{J}_{5}^{\mu}(p,X)$
obtained from chiral kinetic theory up to $\mathcal{O}(\hbar)$ \citep{Hidaka:2018mel}
into the spin polarization vector $\mathcal{S}^{\mu}$, we derive
the helicity polarization \citep{Yi:2021ryh, Yi:2021unq, Wu:2022mkr},
\begin{eqnarray}
S_{\textrm{hydro}}^{h}(\mathbf{p}) & = & S_{\textrm{thermal}}^{h}(\mathbf{p})+S_{\textrm{shear}}^{h}(\mathbf{p})+S_{\textrm{accT}}^{h}(\mathbf{p})\nonumber \\
 &  & +S_{\textrm{chemical}}^{h}(\mathbf{p}),\label{eq:concentrate_on}
\end{eqnarray}
where
\begin{eqnarray}
S_{\textrm{thermal}}^{h}(\mathbf{p}) & = & \int d\Sigma^{\sigma}F_{\sigma}p_{0}\epsilon^{0ijk}\widehat{p}_{i}\partial_{j}\left(\frac{u_{k}}{T}\right),\nonumber \\
S_{\textrm{shear}}^{h}(\mathbf{p}) & = & -\int d\Sigma^{\sigma}F_{\sigma}\frac{\epsilon^{0ijk}\widehat{p}^{i}p_{0}}{(u\cdot p)T}(p^{\sigma}\pi_{\sigma j}u_{k}),\nonumber \\
S_{\textrm{accT}}^{h}(\mathbf{p}) & = & \int d\Sigma^{\sigma}F_{\sigma}\frac{\epsilon^{0ijk}\widehat{p}^{i}p_{0}u_{j}}{T}\left[(u\cdot\partial)u_{k}+\frac{\partial_{k}T}{T}\right],\nonumber \\
S_{\textrm{chemical}}^{h}(\mathbf{p}) & = & -2\int d\Sigma^{\sigma}F_{\sigma}\frac{p_{0}\epsilon^{0ijk}\widehat{p}_{i}}{(u\cdot p)}\partial_{j}\left(\frac{\mu}{T}\right)u_{k},\label{eq:helcity_decomp_01}
\end{eqnarray}
stand for the contributions from thermal vorticity, the
shear viscous tensor, the fluid acceleration minus the gradient of
temperature $T$, the gradient of baryon chemical potential $\mu$
over temperature, respectively. Here, we introduce $\pi_{\sigma j}=\partial_{\sigma}u_{j}+\partial_{j}u_{\sigma}-u_{\sigma}(u\cdot\partial)u_{j}$
and $F^{\mu}=\hbar[8m_{\Lambda}\Phi(\mathbf{p})]^{-1}p^{\mu}f_{eq}(1-f_{eq}),$
$\Phi(\mathbf{p})=\int d\Sigma^{\mu}p_{\mu}f_{eq}$. We also assume
that the system reaches the local thermal equilibrium for simplicity,
i.e. we choose $f_{eq}=1/\left[\text{exp}[(p^{\mu}u_{\mu}-\mu)/T]+1\right]$.
For other decomposition of spin vector, we refer to Refs. \citep{Liu:2019krs,Liu:2020dxg,Liu:2021uhn,Fu:2021pok,Becattini:2021iol,Becattini:2021suc}.
For convenience, we further decompose helicity polarization induced
by thermal vorticity $S_{\text{thermal }}^{h}$ into two separate
terms \cite{Yi:2021unq},
\begin{eqnarray}\label{eq:decomp_thermal}
S_{\nabla T}^{h}(\mathbf{p}) & = & \int d\Sigma^{\sigma}F_{\sigma}\frac{p_{0}}{T^{2}}\widehat{\mathbf{p}}\cdot(\mathbf{u}\times\nabla T),\nonumber \\
S_{\omega}^{h}(\mathbf{p}) & = & \int d\Sigma^{\sigma}F_{\sigma}\frac{p_{0}}{T}\widehat{\mathbf{p}}\cdot\boldsymbol{\omega},
\end{eqnarray}
denoting the polarization related to the gradient of temperature,
and caused by the kinetic vorticity $\bm{\omega}=\nabla\times\mathbf{u}$,
respectively. Later on, we will see that the above decomposition can
improve our understanding of helicity polarization. 

Since the electromagnetic fields generating by the collisions decay
rapidly \citep{Deng:2012pc,Roy:2015coa,Roy:2015kma,Pu:2016ayh,Pu:2016bxy,Inghirami:2016iru, Shi:2017cpu,Siddique:2019gqh,Peng:2022cya}
and are negligible at the late stage, we omit the helicity polarization
induced by electromagnetic fields for simplicity. In general, an axial
chemical potential, characterising local parity violation, near the freeze-out hypersurface also contributes
to the helicity polarization \citep{Becattini:2020xbh,  Gao:2021rom}. However, the event-by-event averaged axial density or axial chemical potential is almost vanishing. We therefore exclude these contributions
in the analysis of helicity polarization. 
We also notice that in the absence of the axial charge the system
has the space reversal symmetry, which leads to $S_{\textrm{hydro}}^{h}(Y,\phi_{p})=-S_{\textrm{hydro}}^{h}(-Y,\phi_{p}+\pi).$
More discussions on the properties of helicity polarization can be
found in Refs. \citep{Becattini:2020xbh,Gao:2021rom,Yi:2021unq}. 

Similar to the local spin polarization in heavy ion collisions, e.g.
see Refs. \citep{Fu:2020oxj,Becattini:2021iol,Ryu:2021lnx,Yi:2021unq,Wu:2022mkr},
we propose a possible physical observable followed Ref. \citep{Yi:2021unq}, 
\begin{eqnarray}
P_{H}(\phi_{p}) & = & \frac{2\int_{Y_{\text{min}}}^{Y_{\text{max}}}dY\int_{p_{T\text{min}}}^{p_{T\text{max}}}p_{T}dp_{T}[\Phi(\mathbf{p})S_{\textrm{hydro}}^{h}]}{\int_{Y_{\text{min}}}^{Y_{\text{max}}}dY\int_{p_{T\text{min}}}^{p_{T\text{max}}}p_{T}dp_{T}\Phi(\mathbf{p})}.\label{eq:def_PH}
\end{eqnarray}
The prefactor $2$ in numerator comes from experimental measurement
of local spin polarization, which is proportional to $\frac{1}{s}\mathcal{S}^{\mu}(\mathbf{p})$
with $s=\frac{1}{2}$ for the $\Lambda$ hyperons. 
Note that the $S^h$ defined in Eq.~(\ref{eq:def_Sh}) is not a Lorentz scalar. In the current study, we define the polarization three-vector
$\boldsymbol{\mathcal{S}}$ in Eq.~(\ref{eq:def_PH}) in
the laboratory frame, while the one defined in the rest frame of $\Lambda$
hyperons can easily be derived by the Lorentz transformation, $\boldsymbol{\mathcal{S}}^{\prime}=\boldsymbol{\mathcal{S}}-\frac{(\mathbf{p}\cdot\boldsymbol{\mathcal{S}})\mathbf{p}}{E_{\Lambda}(E_{\Lambda}+m_{\Lambda})}$
with $E_{\Lambda}$ being the energy of $\Lambda$ hyperons. Similarly,
the helicity polarization can also defined by $\boldsymbol{\mathcal{S}}^{\prime}$,
i.e. $\mathbf{\ensuremath{\widehat{\mathbf{p}}}}\cdot\boldsymbol{\mathcal{S}}^{\prime}=\frac{m_{\Lambda}}{E_{\Lambda}}\mathbf{\ensuremath{\widehat{\mathbf{p}}}}\cdot\boldsymbol{\mathcal{S}}=\frac{m_{\Lambda}}{E_{\Lambda}}S^{h}$.

To investigate
the helicity polarization originating from various sources, we utilize
Eq.~(\ref{eq:concentrate_on}) to express $P_{H}$ as the sum of four
terms and further decompose the thermal-vorticity contribution in light of Eq.~(\ref{eq:decomp_thermal}),
\begin{eqnarray}
P_{H}^{\textrm{total}} & = & P_{H}^{\text{thermal}}+P_{H}^{\text{shear}}+P_{H}^{\text{accT}}+P_{H}^{\text{chemical}},\label{eq:All_effect}\\
P_{H}^{\text{thermal}} & = & P_{H}^{\omega}+P_{H}^{\nabla T},\label{eq:def_decompose_th}
\end{eqnarray}
where the superscripts indicate the respective sources. 

We utilize the (3+1)D CLVisc hydrodynamics \citep{Pang:2012he, Wu:2021fjf, Wu:2022mkr}
to simulate the evolution of the QGP at different collision energies.
We employ the SMASH model \citep{Weil:2016zrk, Schafer:2019edr, Mohs:2019iee, Hammelmann:2019vwd, Mohs:2020awg, Schafer:2021csj, Inghirami:2022afu}
for initial conditions and adopt the NEOS-BQS equations of state \citep{Monnai:2019hkn,Monnai:2021kgu}.
Later on, we will also check the results from the AMPT initial model
\citep{Lin:2004en, Wu:2021hkv, Wu:2018cpc, Zhao:2017yhj}. We have
chosen the simulation parameters according to the studies \cite{Wu:2021fjf,  Wu:2022mkr},
where the hydrodynamic model has been shown to successfully reproduce
the pseudo-rapidity distribution of charged hadrons, as well as the
transverse momentum spectra of protons, pions, and kaons measured
in the experiments. We would like to emphasize that the final results
for the spectra of mesons in low energy collisions are insensitive to a parameter
$C_{B}$, which connects to a baryon diffusion coefficient $\kappa_{B}=\frac{C_{B}}{T}n\left[\frac{1}{3}\cot\left(\frac{\mu}{T}\right)-\frac{nT}{w}\right]$
with $n$ being baryon number and $w$ being enthalpy density \cite{Wu:2021fjf,  Wu:2022mkr}.
We set $C_{B}=0$ in most of our simulations unless explicitly stated
otherwise. From the simulations, we obtain the profile of temperature,
chemical potential and fluid velocity at the chemical freeze-out hypersurface.
By inputting these quantities into Eqs.~(\ref{eq:Cooper-Frye formula},
\ref{eq:concentrate_on}, \ref{eq:def_PH}), we derive the helicity
polarization as a function of the azimuthal angle $\phi_{p}$. The
integration bounds for Eq.~(\ref{eq:def_PH}) is chosen as $p_{T}\in[0.5,3]$ GeV
and $Y\in[-1,1]$. We take the mass of $\Lambda$ hyperons, $m_{\Lambda}=1.116$ GeV
in Eq.~(\ref{eq:Cooper-Frye formula}), as well as the mass term
in all $(u\cdot p)$ terms of Eq.~(\ref{eq:helcity_decomp_01}). 

\begin{figure*}[thb]
\includegraphics[scale=0.35]{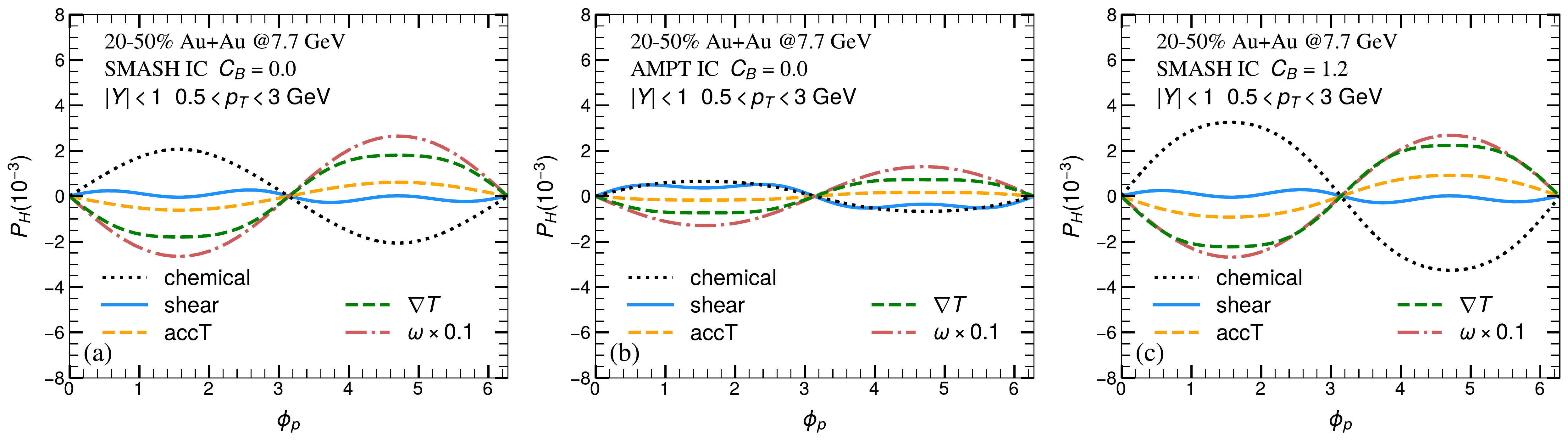}
\caption{(a) Helicity polarization induced by various sources in $20-50\%$
centrality at $\sqrt{s_{NN}}=7.7$ GeV Au+Au collisions with SMASH
initial condition. (b) The results were obtained using the same parameters
as in Fig.~\ref{fig:all_decomposition}(a) except for the initial condition
given by AMPT model. (c) The results are obtained using the same parameters
as in Fig.~\ref{fig:all_decomposition}(a) except for $C_{B}=1.2$.
The shortened form "chemical", "shear", "accT", "$\nabla T$" and
"$\omega\times0.1$" stand for the $P_{H}^{\text{chemical}},P_{H}^{\text{shear}},P_{H}^{\text{accT}},P_{H}^{\nabla T}$
and $0.1\times P_{H}^{\omega}$, respectively. \label{fig:all_decomposition}}
\end{figure*}

\begin{figure}
\includegraphics[scale=0.35]{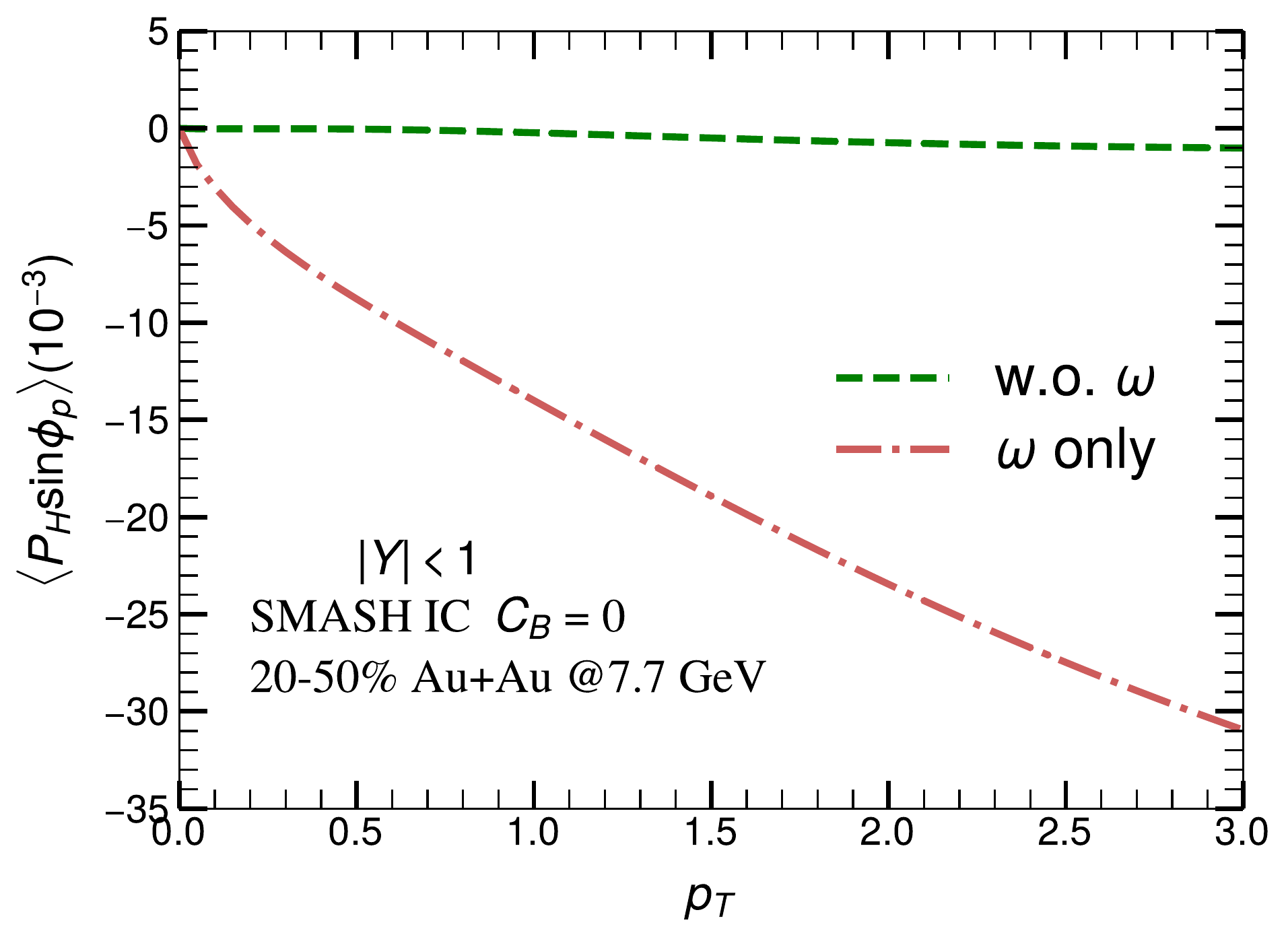}
\caption{The first Fourier sine coefficient of the helicity polarization as
a function of $p_{T}$ in $20-50\%$ centrality at $\sqrt{s_{NN}}=7.7$
GeV Au+Au collisions with SMASH initial condition. The same color
assignments as in Fig. \ref{fig:cb0}. \label{fig:pt}}
\end{figure}

\section{Numerical results from hydrodynamics approaches\label{sec:Numerical-results-from}}

We present the numerical results for helicity polarization $P_{H}(\phi_{p})$
induced by various sources as a function of azimuthal angle in $20-50\%$
centrality at $\sqrt{s_{NN}}=7.7,19.6,39\text{ GeV}$ Au+Au collisions.
Similar to Refs. \citep{Niida:2018hfw,Adam:2019srw,ALICE:2021pzu},
we also compute the first Fourier sine coefficient of the helicity
polarization of $\Lambda$ hyperons, $\langle P_{H}\sin\phi_{p}\rangle$,
as a function of transverse momentum. 


Let us start by discussing the impact of collision energy on helicity
polarization. As the collision energy decreases, nuclear stopping effects
become more prominent and the larger portion of orbital angular momenta of colliding nuclei is transferred to the remnants.  
This leads to an expected growth of the
kinetic vorticity with decreasing collision energies, as has been
suggested by previous studies \citep{Deng:2016gyh,Wei:2018zfb,Deng:2020ygd,Huang:2020xyr}.
The experimental measurement of global polarization agrees with this
expectation; that is, the global polarization increases as the collision
energy decreases \citep{STAR:2017ckg}. Naturally, one may expect that
the helicity polarization induced by kinetic vorticity also follows the same trend, which is clearly observed in Fig.~
\ref{fig:cb0}. 


Furthermore, another important observation from Fig.~\ref{fig:cb0}
is that the $P_{H}^{\omega}$ dominates at intermediate and low collision
energies. 
Unlike the global spin polarization successfully mainly described by the contribution from thermal vorticity in global equilibrium  satisfying the Killing condition \citep{Becattini:2012tc},  
other corrections such as the shear viscous tensor, gradient of baryon chemical potential over temperature should be incorporated in local thermal equilibrium, which play
an important role to the local spin polarization \citep{Fu:2021pok,Becattini:2021iol,Ryu:2021lnx,Yi:2021ryh,Fu:2022myl,Fu:2022oup,Alzhrani:2022dpi,Wu:2022mkr}. 
Based on our previous studies \citep{Wu:2022mkr},  
the magnitude of local spin polarization
along the out-of-plane direction for $\Lambda$ hyperons induced by
shear viscous tensor, fluid acceleration, and $\nabla(\mu/T)$ at
intermediate and low collision energies is much smaller than that
caused by thermal vorticity.
This conclusion holds if we choose the mass of particles
in Eqs.~(\ref{eq:Cooper-Frye formula}) and (\ref{eq:helcity_decomp_01})
as the mass for $\Lambda$ hyperons. Naturally, we anticipate that
the helicity polarization induced by the thermal or kinetic vorticity
dominates over other contributions. 

As an example, we examine the behavior of $P_{H}$ in
$\sqrt{s_{NN}}=7.7$ GeV Au+Au collisions to demonstrate the contributions
from various sources, as shown in Fig.~\ref{fig:all_decomposition}(a).
We observe that the magnitude of $P_{H}^{\omega}$ is approximately
$10$ times greater than that induced by other sources, which is consistent
with the above analysis for the corrections out of global equilibrium. Furthermore,
we observe that the dependence of $\{P_{H}^{\text{chemical}},P_{H}^{\text{shear}}\}$
or $\{P_{H}^{\text{accT}},P_{H}^{\nabla T}\}$ on $\phi_{p}$ resembles
that of the sine or negative sine function, respectively. Eventually,
$P_{H}^{\text{chemical}},P_{H}^{\text{shear}},P_{H}^{\text{accT}},P_{H}^{\nabla T}$
nearly cancel each other out, highlighting the dominant role of kinetic
vorticity in helicity polarization.

A natural question that arises is whether the dominant role of kinetic
vorticity in helicity polarization persists for different initial
conditions or parameters. To verify it, we have studied the helicity
polarization at $7.7$GeV Au+Au collisions as an example using the
AMPT initial conditions in Fig. \ref{fig:all_decomposition}(b), and
with a different baryon diffusion coefficient $C_{B}=1.2$ \cite{Wu:2021fjf,  Wu:2022mkr} 
in Fig. \ref{fig:all_decomposition}(c). 
We find that, regardless of the implemented
initial conditions or the value of the baryon diffusion coefficient
$C_{B}$, $P_{H}^{\omega}$ is always significantly larger than
the helicity polarization induced by other sources even though the magnitudes of helicity polarization from all sources including kinetic vorticity are together enhanced by a nonzero $C_{B}$. Therefore, our
conclusion that $P_{H}^{\omega}$ dominates in $P_{H}^{\textrm{total}}$
is independent of initial conditions or $C_{B}$. The helicity
polarization induced by other sources, excluding the kinetic vorticity,
also approximately cancel out with each other.


Another possible relevant observable in experiments is the first Fourier
sine coefficient of the helicity polarization of $\Lambda$ hyperons
\citep{Niida:2018hfw,Adam:2019srw,ALICE:2021pzu}. We plot $\langle P_{H}\sin\phi_{p}\rangle$
as a function of $p_{T}$ in $20-50\%$ centrality at $\sqrt{s_{NN}}=7.7\text{ GeV}$
as an example in Fig.~\ref{fig:pt}. We find that the magnitude of
$\langle P_{H}\sin\phi_{p}\rangle$ increases with growing $p_{T}$.
Moreover, we observe that helicity polarization induced by kinetic
vorticity still dominates over other contributions, which is consistent with
the results in Fig.~\ref{fig:cb0}.

As a remark, it is noteworthy that $P_{H}^{\omega}$ dominates the
helicity polarization, especially in low energy collisions. Therefore,
by mapping the hydrodynamic simulations to the helicity polarization
measured in future experiments, one can extract the structure of kinetic
vorticity. This provides us with a new opportunity to probe the structure
of kinetic vorticity through helicity polarization.

Before ending this section, we would like to discuss the potential
impact of two crucial approximations for the local spin polarization, namely
strange memory scenario \citep{Fu:2021pok} and isothermal equilibrium
\citep{Becattini:2021iol}, on the helicity polarization. These two
approximations are of great importance in delineating the local spin polarization
of $\Lambda$ hyperons at $\sqrt{s_{NN}}=200$ GeV Au+Au collisions.
However, in low-energy collisions, it is unclear whether the quark
degrees of freedom are released from the hadrons in the fireball.
Therefore, it is plausible to consider the helicity polarization of
$\Lambda$ hyperons rather than $s$ quarks. We have also checked
numerically that in strange memory scenarios, the helicity polarization
induced by other effects, excluding kinetic vorticity, remains negligible
and contributes only to a small percentage of the total helicity polarization.
The sign of $\langle P_{H}\sin\phi_{p}\rangle$ remains unchanged
in the strange memory scenarios. In the isothermal equilibrium, the
temperature gradient near the chemical freeze-out hypersurface is
assumed to be vanishing. We have numerically checked that even if
we drop contributions from temperature gradient, the dominance of
$P_{H}^{\omega}$ in the total helicity polarization still holds. 

\section{Summary \label{sec:Conclusion-and-discussion}}

We have studied the helicity polarization of $\Lambda$ hyperons and
its first Fourier sine coefficient at RHIC-BES energies and observed
that the helicity polarization induced by kinetic vorticity $P_{H}^{\omega}$
dominates at intermediate and low collision energies. The helicity
polarization led by other sources is one order of magnitude smaller
than $P_{H}^{\omega}$ and their net contributions approximately cancel out.
Furthermore, the dominance of $P_{H}^{\omega}$ remains unchanged
by variations in initial conditions and baryon diffusion coefficient. Such a hierarchy for helicity polarization is also unchanged even when adopting the approximations of the strange memory scenarios and isothermal equilibrium in low-energy collisions. 

Based on our results, we propose a novel approach to probe the fine structure of kinetic vorticity by linking hydrodynamic simulations
to the measurements of helicity polarization in future low-energy nuclear collision experiments. On the other hand, at low-energy collisions, the helicity polarization provides a robust baseline for the equilibrium contribution to spin polarization. The sizable mismatch for comparing our predictions with future experimental measurements could reveal the potential role of non-equilibrium contributions from collisional effects or even more exotic sources to local spin polarization.   

\section*{Acknowledgments}
This work is supported
in part by the National Key Research and Development Program of China
under Contract No. 2022YFA1605500 and is also supported by National
Natural Science Foundation of China (NSFC) under Grants No. 12075235,
12135011, 12225503, 12175123, 11890710, 11890711, 11890713, 11935007 and 11475104. D.-L. Y. was supported by National Science and Technology Council (Taiwan) under Grant No. MOST 110-2112-M-001-070-MY3.

\bibliographystyle{h-physrev}
\bibliography{qkt-ref20230407}

\end{document}